\documentclass[manuscript]{aastex}

\begin{document}
\pagenumbering{arabic}

\title{SHAPLEY-AMES GALAXIES IN THE BLUE AND INFRARED}

\author{Sidney van den Bergh}
\affil{Dominion Astrophysical Observatory, Herzberg Institute of Astrophysics, National Research Council of Canada, 5071 West Saanich Road, Victoria, BC, V9E 2E7, Canada}
\email{sidney.vandenbergh@nrc.gc.ca}

\begin{abstract}

   The Shapley-Ames Catalog of 1276 galaxies with B $<$ 12.5 is 
compared with the Sanders et al. all sky sample of the 629 galaxies 
with 60 $\mu$m flux density \linebreak
$>$ 5.24 Jy. The fraction of Shapley-Ames 
galaxies that are visible in the IR is found to increase 
from 0.006 for E and E/S0 galaxies to 0.384 for Sc galaxies. 
The subset of Shapley-Ames galaxies that are detected in the 
IR has a median blue luminosity that is $\sim$0.8 mag fainter than 
that of all Shapley-Ames galaxies. Most of this difference is 
due to the fact that late-type galaxies (which contain 
dust and hot stars) are systematically less luminous in blue 
light than are early-type galaxies. Within individual stages 
along the Hubble sequence no significant differences are found 
between the luminosity distributions in blue light of galaxies 
that were detected in the infrared and those that were not. 
However, our data show a puzzling exception (significant at 
99.9\% ) for SBc galaxies. For reasons that are not understood 
Shapley-Ames SBc galaxies, that are visible in the IR, are 
more luminous in blue light than those SBc galaxies that are not 
detected in the infrared. An other peculiarity of the data is that 
Shapley-Ames Sc galaxies galaxies are (at 99.6\% confidence) 
more luminous in blue light than objects of type SBc.

\end{abstract}


\section{INTRODUCTION}

   Two decades ago Sandage \& Tammann (1981) published {\it A Revised 
Shapley-Ames Catalog of Bright Galaxies} which contained photometry, 
and uniform morphological classifications, for 1276 galaxies with 
total apparent blue magnitudes brighter than B = 12.5, although the 
data become increasingly incomplete for B $>$ 12.0. These observations 
may be compared to recent complete all sky infrared photometry of 
629 galaxies with a total 60 $\mu$m flux density greater than 5.24 Jy by 
Sanders et al. (2003). Such a comparison of the Shapley-Ames Catalog 
with the data by Sanders et al. shows that 298 of these galaxies 
have both B $<$ 12.5 mag and 60 $\mu$m flux $>$ 5.24 Jy. It is the purpose 
of the present note to see how this sub sample of 298 galaxies 
visible in both the blue and in the infrared differs from the entire 
galaxy sample in the Shapley-Ames Catalog. The present work may 
be regarded as an update of a similar study by de Jong et al. (1984), 
which covered only 165 galaxies observed at 60 $\mu$m. It would be 
particularly interesting to test their conclusion that the 
rate of star formation in barred spirals is greater than that in 
normal spirals. The more extensive data that are now 
available should also allow one to firm up the conclusion by de Jong et al. 
that the fraction of infrared emitting galaxies depends on Hubble type 
in the sense that spiral galaxies are more likely to be infrared emitters 
than are ellipticals.  All galaxy classifications and luminosities used in the present paper were drawn from  {\it A Revised Shapley-Ames Catalog} (Sandage \& Tammann 
1981).  In doing the statistics a galaxy classified as say Sb/Sc was counted 
as 0.5 Sb and 0.5 Sc. 

\section{MORPHOLOGY DISTRIBUTIONS OF GALAXIES}

   Table 1 shows a comparison between the frequency distributions 
of various galaxy types in the Shapley-Ames Catalog (Sandage \& 
Tammann 1981, p. 91) [which are denoted SA] and the distribution 
of morphological types in the subset of Shapley-Ames galaxies that 
were detected in the infrared by Sanders et al. (2003) [which is 
denoted by IR]. The most striking feature of these data (see 
Table 2)  is that the fraction of galaxies that are detected in the 
infrared is a strong function of morphological type. Such an effect 
is expected because the 60 $\mu$m radiation from galaxies is dominated 
by emission from warm dust surrounding young hot stars. Only 0.6\% of 
E + E/S0 galaxies in the Shapley-Ames Catalog were seen in the IR 
compared to 38.4\% of Shapley-Ames  Sc galaxies. [The single E galaxy 
in the sample that was detected in the IR is NGC 1275, which may 
actually represent the collision of a spiral and a giant elliptical 
(Baade \& Minkowski 1954)]. 

A comparison between all Shapley-Ames galaxies, and the sub sample 
of those galaxies that were detected at 60 $\mu$m, shows that the median 
blue luminosity of the IR sub sample is $\sim$0.8 mag fainter than the median 
luminosity of all Shapley-Ames galaxies. The main reason for this 
difference is that late-type galaxies (which contain hot young stars that are able to heat dust) are systematically less luminous in blue light [see Figure 1 
of van den Bergh (1997)] than early-type galaxies. However, the present 
data suggest that the truth might actually be a bit more subtle than this. 
Kolmogorov-Smirnov tests show no significant difference between the 
luminosity distributions in blue light of Sa and SBa, Sab and SBab, Sb and 
SBb, Sbc and SBbc galaxies that are (or that are not) detected in the IR. 
However, a K-S test does show that SBc galaxies that are visible in the 
infrared are (at the 99.7\% significance level) more luminous in blue 
light than are those that were not detected in the IR. This result may also be seen in a different way by considering the contingency tables shown as Tables 
3 and 4. Table 3 shows such a 2x2 contingency table in which Sc galaxies are split 
into objects that where detected (or not detected) in the IR and that are 
brighter (or fainter) than $M _{B} = -21.0$. A similar table for SBc galaxies is presented  in Table 4. The latter contingency table yields a probability of less than 0.1\% that the bright and faint SBc galaxies that are IR-bright were drawn from the same parent population of blue luminosities as those that 
are IR-dark.  The observed difference is in the sense that IR-bright SBc 
galaxies are more luminous than are IR-dark SBc spirals.  (Little or no
difference is seen between the IR bright and IR dark Sc galaxies in 
Table 3). The reason for the observed difference between IR bright and IR dark 
SBc galaxies is presently not understood. 

The data in Table 5 show that 29\% of all normal 
spirals were detected in the infrared, compared to only 21\% of barred spirals. 
The data in this contingency table show that the difference in IR 
detection probability between  S and SB galaxies is significant at the 99\% 
level. This difference in the IR delectability of S and SB galaxies is almost 
entirely due to the fact that faint Sc galaxies are so much more likely to be 
detected in the IR than is  the case for faint SBc galaxies. This difference 
is puzzling because the Sc and the SBc galaxies fainter than $M_{B}= -21.0$
have almost exactly the same mean luminosities in blue light. The 
difference in their IR properties  might be due to either (1) faint Sc galaxies 
being dustier than faint SBc galaxies, or (2) to the dust in faint Sc galaxies 
being closer to the hot illuminating stars than is the case in galaxies of type 
SBc. Neither of these two alternatives  appears very attractive. 

\section{FLUX DEPENDENCE OF GALAXY CHARACTERISTICS}

The bi-variate luminosity function of galaxies (cf. Auriemma et al. 1977) 
can sometimes provide additional information on galactic evolution. However, 
a subdivision of data samples only provides additional insights for large samples. 
In the present case the sample of 298 Shapley-Ames galaxies in the infrared catalog 
of Sanders et al. (2003) may be divided into an IR bright sub sample  of 157 galaxies with 
60 $\mu$m  flux $\geq$ 10.0 Jy, and an IR faint sub sample with  141 objects having 10.00 
Jy $< f < $ 5.24 Jy. Inspection of the data (and comparison with Table 1) appears to show no 
obvious differences between the distribution of the IR bright and IR faint sub samples 
over Hubble type. Table 6 shows the distribution of 60 $\mu$m IR radiation for Sc galaxies 
as a function of parent galaxy blue luminosity. The table has Chi$^{2}$ = 5.06 (with two 
degrees of freedom) yielding only an 8\% probability that luminous Sc galaxies are 
more likely to be observed in the IR than are less luminous Sc galaxies. In this  this 
respect Sc galaxies differ significantly from those of type SBc for which a strong 
dependence of IR luminosity on blue luminosity was found in Section 2. Table 7 
clearly shows that this effect is entirely due to the striking difference in IR detections 
between  bright $(M _{B} \leq -21.0)$ and faint $(M _{B} > -21.0)$ SBc galaxies. The reason 
for this enormous difference is not clear. 

 A comparison between the luminosity distributions of Shapley-Ames Sc and 
SBc galaxies shows that there are, relatively speaking, fewer faint Sc galaxies 
than faint SBc galaxies. A Kolmogorov-Smirnov test shows that this  excess of 
faint SBc galaxies is significant at the 99.6\% level. Posssibly this apparent difference 
in the luminosity distributions of  Sc and SBc is due to a small systematic effect in the 
galaxy classifications by Sandage and Tammann. Alternatively very late-type dwarfs 
might be more prone to bar formation than galaxies of slightly earlier morphological 
types. Statistical weak support for the latter suggestion is provided by the observation 
that  8.5 out 20 (42\%)  of the Sm and Im galaxies (which are all intrinsically faint) in the 
Shapley-Ames catalog are barred, compared to only 24\% barred objects among the 
Shapley-Ames galaxies of type Sc with $(M _{B} > -21.0)$.

\section{CONCLUSIONS}

   The luminosity of a galaxy at 60 $\mu$m is determined by (1) the rate of 
star formation, by (2) the amount of dust present in that galaxy and (3) 
by the relative distribution of hot young stars and the dust (Misiriotis 
et al. 2004). One would expect actively star forming late-type galaxies 
to be stronger IR emitters than inactive early-type galaxies. This 
expectation is strongly confirmed by the data in Table 2 which show that the 
fraction of Shapley-Ames galaxies, that are observed in the IR, rises from 
only 0.6\% among E + E/S0 galaxies to 38.4\% for Sc galaxies. It is noted in 
passing that the very low frequency of IR emission among elliptical galaxies also suggests that the rate at which early-type galaxies swallow late-type 
spirals must presently be quiet low. The new data by Sanders et al. 
 (2003) do not support the tentative suggestion by de Jonge et al. (1984) 
 that the rate of star formation in barred spirals might be higher than 
that in normal spirals. In fact the more modern data suggest (at 
99\%  confidence) that normal spirals are actually more likely to be 
visible in the IR than is the case for barred spirals. For all galaxy types 
(except SBc) the distributions of blue luminosities for all objects 
 that are detected in the IR is statistically indistinguishable from that 
 of those that are not detected in the infrared. However, a puzzling 
 exception is provided by SBc galaxies that are unexpectedly faint in blue light.
 It turns out that such objects are much less likely to be IR emitters 
than is the case for Sc galaxies that are luminous in blue light. 
No reasonable explanation is found for this effect. 

I should like to thanks Colin Borys and Doug Johnstone for their 
kind help with references to the literature on dust and infrared 
emission.   Finally I express my thanks to the referee, Prof. T. de Jong, for impressing on me the importance of the bi-variate luminosity function of galaxies.

 

\begin{deluxetable}{lrrlrr}
\tablewidth{0pt}        
\tablecaption{Frequency distribution of Hubble types}
\tablehead {Normal spirals & & & Barred spirals & &\\
Type & IR & SA & Type & IR & SA}
\startdata
E + E/S0   & $1$      &  $173$    & & &                                    \\
S0 + S0/a  & $6.5$    &  $142$    &    SB0 + SB0/SBa  &     $1$     &  $48$   \\
Sa + Sab   & $23.5$   &  $123$    &    SBa + SBab     &     $6.5$   &  $42$    \\
Sb + Sbc   & $71.5$   &  $187$    &    SBb + SBbc     &     $30.5$  &  $96$    \\
Sc         & $112.5$  &  $293$    &    SBc            &     $19$    &  $77$    \\
Scd + Sd   & $4$      &  $26$     &    SBcd + SBd     &     $0$     &  $8$     \\
Sm  + Im   & $7$      &  $13$     &    SBm + IBm      &     $2$     &  $9$     \\
S          & $4$      &  $16$     &    SB             &     $0$     &  $5$     \\
Special    & $8$      &  $18$     &                   &             &          \\
Total      & $238$    &  $991$     &                   &     $59$    &  $285$   \\
\enddata
\end{deluxetable}

\begin{deluxetable}{lr}
\tablewidth{0pt}        
\tablecaption{Percentage of normal spirals detected at 60 $\mu$m}
\tablehead {\colhead{Type}   & \colhead{IR detections}}
\startdata

E + E/S0     &    $1\%$    \\ 
S0 + S0/a    &    $5\%$    \\
Sa + Sab     &    $19\%$    \\
Sb + Sbc     &    $38\%$    \\
Sc           &    $38\%$     \\

\enddata
\end{deluxetable}

\begin{deluxetable}{lll} 
\tablewidth{0pt}          
\tablecaption{IR detection frequency for luminous and faint Sc galaxies} 
\tablehead {\colhead{Luminosity}      &    \colhead{IR detected}  &   \colhead{Not seen in IR}}
\startdata
$M_{B} \leq -21.0$                    &       47                 &      74.5      \\
$M_{B} >  -21.0$                      &      ~59.5               &     108        \\

\enddata
\end{deluxetable}

 
\begin{deluxetable}{lrr}
\tablewidth{0pt}    
\tablecaption{IR detection frequency for luminous and for faint SBc galaxies}
\tablehead {\colhead{Luminosity}   & \colhead{IR detected}    &    \colhead{Not seen in IR}}
\startdata
$M_{B} \leq -21.0$                  &       12.5                 &       8.5           \\
$M_{B}  >   -21.0$                  &        6.5                 &      47.5           \\
        
\enddata
\end{deluxetable}

\begin{deluxetable}{llr}
\tablewidth{0pt}    
\tablecaption{IR detection frequency for normal and barred spiral \tablenotemark{a}}. 
\tablehead{\colhead{Type}           & \colhead{S}              & \colhead{SB}}
\startdata         
Detected in IR                      &      237                 &    59   \\
Not detected in IR                  &      581                 &   226   \\  

\tablenotetext{a}{E and E/S0 galaxies excluded}
 
\enddata
\end{deluxetable}

\begin{deluxetable}{lccc}
\tablewidth{0pt}    
\tablecaption{Correlation between 60 $\mu$m flux {\it f} and optical luminosity of Shapley-Ames for galaxies of type Sc}. 
\tablehead{\colhead{Blue luminosity} & \colhead{{\it f} $\geq$ 10.00 Jy} & \colhead{10.00 $> f >$ 5.24 Jy}  & \colhead{$f <$5.24 Jy}}
\startdata  
       
$(M _{B} \leq -21.0)$     & 31     &     16      &     ~~74.5\\            
$(M _{B} > -21.0)$        &~~27.5    &    32      &     108 \\

\enddata
\end{deluxetable}

\begin{deluxetable}{lccc}
\tablewidth{0pt}    
\tablecaption{Correlation between 60 $\mu$m flux {\it f} and optical luminosity of Shapley-Ames for galaxies of type SBc}
\tablehead{\colhead{Blue luminosity} & \colhead{$f \geq$ 10.00 Jy} & \colhead{10.00 $> f >$ 5.24 Jy}  & \colhead{$f <$5.24 Jy}}
\startdata  
       
$(M _{B} \leq -21.0)$    &  6     &   ~~6.5      &     8.5  \\            
$(M _{B} > -21.0)$       &  ~~3.5   &    3        &     47.5  \\

\enddata
\end{deluxetable}

  
\end{document}